\documentclass[a4paper,11pt]{article}
\pdfoutput=1 

\usepackage{jheppub} 

\usepackage[T1]{fontenc} 
\usepackage{commands}
\usepackage{dsfont}

\renewcommand{\e}{e}
\renewcommand{\ln}{\log}

\title{'t Hooft suppression and holographic entropy}

\author{William R. Kelly,}
\author{Kevin Kuns,}
\author{and Donald Marolf}

\affiliation{University of California, Santa Barbara \\ Santa Barbara, CA 93106, USA}

\emailAdd{wkelly@physics.ucsb.edu}
\emailAdd{kuns@physics.ucsb.edu}
\emailAdd{marolf@physics.ucsb.edu}

\abstract{Recent works have related the bulk first law of black hole mechanics to the first law of entanglement in a dual CFT.  These are first order relations, and receive corrections for finite changes.  In particular, the latter is naively expected to be accurate only for small changes in the quantum state.  But when Newton's constant is small relative to the AdS scale, the former holds to good approximation even for classical perturbations that contain many quanta.  This suggests that -- for appropriate states -- corrections to the first law of entanglement are suppressed by powers of $N$ in CFTs whose correlators satisfy 't Hooft large-$N$ power counting.  We take first steps toward verifying that this is so by studying the large-$N$ structure of the entropy of spatial regions for a class of CFT states motivated by those created from the vacuum by acting with real-time single-trace sources.  We show that $1/N$ counting matches bulk predictions, though we require the effect of the source on the modular hamiltonian to be non-singular.  The magnitude of our sources is $\epsilon N$ with $\epsilon$ fixed-but-small as $N\rightarrow \infty$.    Our results also provide a perturbative derivation -- without relying on the replica trick -- of the subleading Faulkner-Lewkowycz-Maldacena correction to the Ryu-Takayagi and Hubeny-Rangamani-Takayanagi conjectures at all orders in $1/N$. }

\begin{document}
\maketitle
\flushbottom

\section{Introduction} \label{sec:intro}

There has been much activity exploring the intriguing connection between entanglement in holographic field theories and the gravitational field equations of the bulk dual.  This program traces its roots to Jacobson's seminal paper~\cite{Jacobson:1995ab}, which proposed the Einstein equation to be a thermodynamic equation of state for some unknown quantum mechanical system in which the area of surfaces measures entanglement entropy across causal horizons.  Several groups~\cite{Lashkari:2013koa,Faulkner:2013ica,Swingle:2014uza,Faulkner:2014jva} have now used related arguments to derive the linearized gravitational field equations in the context of the anti-de Sitter/conformal field theory (AdS/CFT) correspondence, where the underlying quantum system is well understood; see also the related works~\cite{Nozaki:2013vta,Bhattacharya:2013bna,Jacobson:2015hqa}.   A key assumption in these more recent derivations is the Ryu-Takayangi (RT) conjecture~\cite{Ryu:2006bv,Ryu:2006ef} and its covariant generalization by Hubeny-Rangamani-Takayanagi (HRT) \cite{Hubeny:2007xt}, both of which relate entanglement entropy of subregions of the field theory to the geometry of bulk surfaces.  A partial converse of this result -- that the bulk field equations imply aspects of the Ryu-Takayanagi conjecture -- has also been argued by Lewkowycz and Maldaena~\cite{Lewkowycz:2013nqa}, though see \cite{Haehl:2014zoa} for a discussion of the so-called homology constraint,  \cite{Fischetti:2014zja} for questions about the possible role of complex bulk surfaces, and \cite{Fischetti:2014uxa} for concerns regarding strong time-dependence.

Much of the recent discussion has centered on the so-called first law of entanglement
\begin{align} \label{eq:FirstLaw}
\delta S_A = \delta\langle H_A\rangle\, .
\end{align}
Here $A$ is a subregion of some Cauchy surface for the CFT, $S_A := - \tr(\rho_A \log(\rho_A))$ is the von Neumann entropy of the associated reduced density matrix $\rho_A$, $H_A := - \log(\rho_A)$ is the modular Hamiltonian, and $\delta$ denotes the first variation with respect to the state when the operator $H_A$ is held fixed on the right hand side.  The relation \eqref{eq:FirstLaw} holds at first order, but receives corrections for finite changes.

According to the HRT conjecture, in holographic theories the left hand side of~\eqref{eq:FirstLaw} can be extracted from the bulk geometry.  The right hand side is typically difficult to evaluate, though it reduces to a simple integral of the CFT stress tensor~\cite{Hislop:1981uh} (see also~\cite{Casini:2011kv}) when $A$ is a ball-shaped region and the system is in its global vacuum state.  The combination of these two results makes~\eqref{eq:FirstLaw} a useful formula for studying the relationship between entanglement and geometry.  In particular, in this context \eqref{eq:FirstLaw} coincides with the first law of black hole mechanics in the bulk applied to the Rindler-like Killing horizon defined by a ball-shaped region $A$ on the AdS boundary \cite{Casini:2011kv}.  This argument for the first law can then be inverted to derive linearized bulk equations of motion from HRT~\cite{Lashkari:2013koa,Faulkner:2013ica,Swingle:2014uza,Faulkner:2014jva}.

The starting point for our work is the observation (see e.g. \cite{Swingle:2014uza}) that the entanglement first law \eqref{eq:FirstLaw} is generally useful only for infinitesimal changes in the quantum state; higher order corrections tend to make significant contributions when it instead undergoes any substantial change.  In contrast, the bulk first law of black hole mechanics accurately describes classical deformations -- typically involving very large numbers of quanta -- so long as the changes in entropy and energy are small in comparison with their background values.  In particular, corrections to the bulk first law stem from gravitational back-reaction and are thus suppressed by powers of the bulk Newton constant $G_N$.  This suggests that corrections to the entanglement first law \eqref{eq:FirstLaw} will be correspondingly suppressed, at least for states that would be appropriately semi-classical with respect to the bulk.  Our goal below is to show suppressions by powers of $N$ in similar computations involving the entropy of spatial regions for CFTs whose correlators satisfy 't Hooft power counting and whose spectrum of light operators is sufficiently sparse.  We consider a class of CFT states motivated by those created from the vacuum by acting with real-time single-trace sources, though we require the source to have non-singular effects on the modular Hamiltonian.  In order to model sources that would produce semi-classical coherent states in any bulk dual, the magnitude of our sources is taken to be $\epsilon N$ with $\epsilon$ fixed-but-small as $N\rightarrow \infty$.  See also \cite{Goykhman:2015sga} for other investigations of entanglement in large $N$ gauge theories.

For such states, we verify bulk predictions of powers of $N$ via a direct calculation in the CFT.    Though it differs in detail, the suppression found here is analogous to the large $N$ suppression of such corrections found previously in \cite{Marolf:2003sq,Marolf:2004et}.  Our analysis also provide additional benefits.  First, our explicit formula for second order relative entropy makes manifest the agreement with appropriately-integrated bulk stress tensors -- here defined to include contributions from the stress tensor of bulk gravitons -- required by comparing corrections to the bulk and boundary first laws \cite{Blanco:2013joa,Lin:2014hva}.  As a result, it again demonstrates that a consistent holographic theory of gravity must couple universally to all forms of bulk stress-energy.  It also provides a perturbative derivation of the Faulkner-Lewkowycz-Maldacena subleading correction to the RT and HRT conjectures at all orders in $1/N$~\cite{Faulkner:2013ana}.

Before proceeding, we should elaborate on the above restriction to sources with non-singular action on the modular Hamiltonian.  As explained in section \ref{sec:assumptions} below, from a dual bulk point of view this requires our perturbations to vanish in some neighborhood of the bifurcation surface of our bulk Rindler-like horizon.  So the leading-order large-$N$ RT or HRT entanglement cannot change.  They do, however, affect the above-mentioned order $N^0$ entanglement.  They also change $\left< H_A \right>$ and thus the relative entropy $R_A = \ev{H_A} - S_A$ at order $N^2$.  We will show that these powers of $N$ are correct at all orders in $\epsilon$; the fact that $\Delta S_A$ is of smaller order in $N$ than $S_A$ itself is the $1/N$ suppression advertised above.

The rest of the paper is organized as follows.  We begin with a brief description of our setup in section~\ref{sec:assumptions}. Section~\ref{sec:computation} then computes $\left< H_A \right>, R_A$ and $S_A$ to second order for our family of states.  It also argues to all orders that the powers of $N$ in $\left< H_A \right>, R_A$ and $S_A$ are precisely those predicted by intuition from a bulk dual.  We conclude in section~\ref{sec:discussion} with a brief discussion of our results as well as comments on possible extensions.  Appendix~\ref{app:RA} derives a simple and completely general formula~\eqref{eq:delta2RA} for the second order change in $R_A$ in bipartite quantum systems that gives the above-mentioned explicit formulas for $R_A$ and $\Delta S_A$.

\section{Setting and assumptions} \label{sec:assumptions} \label{sec:states}

We wish to study excitations of the vacuum $|0\rangle$ of a large $N$ CFT in $d$ spacetime dimensions on $\mathbb{R} \times S^{d-1}$. Since we take our inspiration from a possible bulk dual, we impose assumptions similar to those in e.g.\ \cite{Heemskerk:2009pn} and \cite{Hartman:2013mia,Hartman:2014oaa}, taking our CFT to satisfy 't Hooft large-$N$ factorization \cite{tHooft:1973jz} and to have a sparse spectrum of light operators.  As usual, light operators are those whose scaling dimension $\Delta_i$ of ${\cal O}^i(x)$ remains finite as we take $N\to \infty$ and the sparse spectrum condition requires that for any fixed $\Delta$ the number of such operators with $\Delta_i < \Delta$ remains finite at large $N$.

The factorization condition states that the set of light single-trace gauge-invariant local operators  should admit a basis $\{ {\cal O}^i(x) \}$ for which
\begin{align} \label{eq:tHooftRule}
\left< {\cal O}^{i_1}(x_1) \dots {\cal O}^{i_k}(x_k) \right>_c \sim N^{2-k} \, ,
\end{align}
where $\left<\dots\right>_c$ is the connected vacuum correlator and similar notation without the subscript $c$ will also be employed for the full correlator. Furthermore, heavy operators (those that do not remain light as $N\rightarrow \infty$) decouple in the sense that connected correlators involving both heavy and light operators are much smaller. In fact, we assume that we study a low energy process from which operators of finite-but-large dimension are sufficiently decoupled that  sums over the ${\cal O}^i$ below always converge.  Indeed, our only use of the sparse spectrum condition will be to assume that such sums with coefficients of order $N^p$ converge to a result of the same order in $N$.

It will be convenient to take the basis operators ${\cal O}^i$ to have definite scaling dimension $\Delta_i$ at large $N$ and in fact to diagonalize the order $N^0$ term in the connected two-point function.  We also require
\begin{align}
\label{eq:vev}
\left< {\cal O}^i(x) \right> = 0 \, ,
\end{align}
for all $i$, which can be achieved by subtracting appropriate expectation values. Note that we have not required the ${\cal O}^i$ to be scalars; we have merely suppressed any tensor or spinor indices.  Thus, up to the above subtractions, one member (say, ${\cal O}^0$) of our basis  is $1/N$ times the CFT stress tensor which necessarily satisfies $\Delta =d$; i.e. ${\cal O}^0 = \frac{1}{N} (T - \langle T \rangle)$, where we have suppressed the spacetime indices $ab$.

Our expectations that corrections to the first law of entanglement are suppressed arise from considering the semi-classical behavior of a supposed bulk gravitational dual.  Semi-classical bulk states can be created from the vacuum through the action of large classical sources for the perturbative bulk fields.  As usual, we may choose to locate these sources at the boundary where they may be translated into sources for the local single-trace ${\cal O}^i$ above.  Now, single and multi-trace sources mix under time-evolution, but this mixing is again controlled by the $1/N$ expansion:   since the stress tensor generates time evolution, to any order in $1/N$ a light operator ${\cal O}(x)$ can be replaced inside such correlators with an operator at another time that is a sum over $k$ of $k$-trace terms weighted by $1/N^{k-1}$.  Semi-classical behavior is preserved in time, so it should suffice to restrict  attention to states of the form
\begin{subequations}
\label{coherent-state}
\begin{gather} \label{Ualpha}
\ket{\alpha} := U  \ket{0 } \, , \qquad  U := T e^{-i \alpha  J}, \\
\label{eq:source1}
J :=   \sum_{k=1} N^{-(k-1)} \int dx_1 \dots dx_k \ j_{i_1\dots i_k} (x_1,\dots,x_k)    {\cal O}^{i_1}(x_1) \dots {\cal O}^{i_k}(x_k)
  \, ,
\end{gather}
\end{subequations}
where the classical sources $j_{i_1\dots i_k}$ are fixed smooth  $c$-number functions of order $N^0$
and we allow distributional terms so that terms in the $k$-trace contributions to \eqref{eq:source1} may effectively include only $m < k$ integrals over operator location.

We have introduced the real number $\alpha$ to be used as an expansion parameter.  The symbol $T$ denotes time-ordering, and we employ the convenient abuse of notation that defines $T e^{-i \alpha  J}$ to be the natural time-ordered exponential associated with the particular representation of $J$ given above as an integral over the CFT spacetime.  Using the standard AdS/CFT dictionary, the above normalizations would give the bulk field $\phi^i$ dual to ${\cal O}^i$ an expectation value $\langle \phi\rangle_\alpha\sim \alpha$.

We wish to choose $\alpha$ so that \eqref{coherent-state} would behave semi-classically in a bulk gravitational dual.  Since the bulk is perturbative at large $N$,
bulk quantum fluctuations become negligible in the limit $\alpha\gg1$ and gravitational back reaction scales like
\begin{align}
G \langle \phi\rangle^2_\alpha \sim \frac{\alpha^2}{N^2} \, .
\end{align}
To allow this effect to be as large as possible consistent with a perturbative treatment we take
\begin{align}
\alpha = \epsilon N \gg 1 \,
\end{align}
where $\epsilon$ is a small parameter to be held constant in the limit $N\to \infty$.

Our particular interest is in the effect of such sources on a region $A$ of some Cauchy surface $\Sigma$ in the CFT spacetime, or more generally on the associated domain of dependence $D(A)$.  The region $A$ is held fixed as we take $N \to \infty$.  We denote the complementary region on this Cauchy surface by $A^c$, so that $\Sigma = A \cup A^c$. As a density matrix, the state $|\alpha\rangle$ is $\sigma_\alpha := \ket{\alpha} \bra{\alpha}$. The associated reduced density matrices and the un-perturbed modular Hamiltonians  for $A$ and $A^c$ are then
\begin{gather}
\label{eq:rhodef}
\rho_{A\alpha} = \rho_A := \tr_{A^c} \lp\sigma_\alpha \rp \, , \qquad H_A := -\log\lp \rho_{A({\alpha =0)}} \rp\, , \cr
\rho_{A^c\alpha} = \rho_{A^c} := \tr_{A} \lp\sigma_\alpha \rp \, , \qquad H_{A^c} := -\log\lp \rho_{A^c(\alpha =0)} \rp\, .
\end{gather}
As implied above, we will often suppress the label $\alpha$ on $\rho_A, \rho_{A^c}$ when the meaning is clear, though $H_A$, $H_{A^c}$ will always represent the modular Hamiltonians at $\alpha=0$.
An important property of these objects is
\begin{equation}
\label{commute}
[H_A, H_{A^c}] =0,
\end{equation}
where both $H_A$ and $H_{A^c}$ are interpreted as operators on the full CFT Hilbert space by tensoring each with the identity operator ($\mathds{1}_{A^c}$ or $\mathds{1}_{A}$) on the complementary region.
The result \eqref{commute} would be immediate for truly bipartite systems that decompose as a tensor product of a system on $A$ with a system on $A^c$.  It is less trivial in quantum field theory, but still holds for ball-shaped regions $A$ in CFTs since the vacuum is invariant under a $Z_2$ conformal symmetry that exchanges $A$ and $A^c$.  This symmetry must then exchange $H_A$ and $H_{A^c}$ as well, requiring their commutator to be symmetric.  The manifest antisymmetry of $[H_A, H_{A^c}]$ then forces the commutator to vanish.

It is also useful to define the operator $K := H_A - H_{A^c}$.  Recall that $K$ annihilates the vacuum:
\begin{align}
\label{eq:H0}
K \ket{0} = (H_A - H_{A^c}) \ket{0} = 0 \, .
\end{align}
For ball-shaped regions $A$ in a constant-time slice, $K$ generates the conformal isometry of $\mathbb{R} \otimes S^{d-1}$ that moves $A$ orthogonally to itself while preserving the domain of dependence $D(A)$; i.e., it generates the natural forward ``time-translation'' on $D(A)$  and the natural backward ``time-translation'' on $D(A^c)$ \cite{Hislop:1981uh}.  For lack of a better name, we will refer to $K$ as the boost Hamiltonian, though this term is only an accurate description of $K$ in the large-sphere limit when $A$ becomes a half-plane.

Finally, we introduce a basis for the space of light local gauge-invariant single trace operators $\{ {\cal O}^i_A(x) \}$ ($\{ {\cal O}^i_{A^c} (x)\}$) on the domain of dependence $D(A)$ ($D(A^c)$) that again satisfy \eqref{eq:tHooftRule} and \eqref{eq:vev}.  Below we will consider ball-shaped regions in a constant-time Cauchy surface $\Sigma = S^{d-1}$ so that the modular Hamiltonians $H_A, H_{A^c}$ can be expressed as integrals of the stress tensor over $A, A^c$ in any CFT \cite{Hislop:1981uh}.

\subsection{Adapting the source}

A key step in our argument will be to write the source-operator $J$ from \eqref{eq:source1} in a manner adapted to the decomposition $\Sigma =  A \cup A^c$.  The basic idea is to use the Heisenberg equations of motion for the CFT to express $J$ in terms of operators on $\Sigma$.  Thinking of the CFT Hilbert space as a tensor product of separate Hilbert spaces for $A$ and $A^c$ would then allow $J$ to be written as a sum of terms, each of which is the tensor product of a (possibly trivial) operator on $A$ with one on $A^c$.
\begin{align} \label{eq:piDecomposition}
J = \sum_a J^a_A \otimes J^a_{A^c} \, .
\end{align}
We would like to then expand the $J^a_A, J^a_{A^c}$ as a sum or products of the $\{ {\cal O}^i_A(x) \}$ ($\{ {\cal O}^i_{A^c} (x)\}$) and use \eqref{eq:tHooftRule} to count powers of $N$.

There are, however, several potential obstacles to consider.  First, evolving the source to the Cauchy surface $\Sigma$ will generally mix the ${\cal O}_i$ with both heavy operators and highly non-local expressions such as Wilson loops that cannot be expanded in terms of the $\{ {\cal O}^i_A(x) \}$ ($\{ {\cal O}^i_{A^c} (x)\}$).  However, our calculations below will involve only correlation functions of sources with $H_A, H_{A^c}$.  For the ball-shaped regions to be considered, up to the choice of zero-point our $H_A, H_{A^c}$ are integrals of the stress tensor, which is one of our light operators.  Since the stress tensor also generates time evolution, to any order in $1/N$ a light operator ${\cal O}(x)$ can be replaced inside such correlators with an appropriately-smeared sum of products of $\{ {\cal O}^i_A(x) \}$ ($\{ {\cal O}^i_{A^c} (x)\}$).  So in this sense we may write
\begin{align} \label{eq:piAtLAdS}
 J = {\cal O}_{A} + {\cal O}_{A^c}
 + \sum_{k=2} \frac{1}{N^{k-1}}  \int_{[D(A)\cup D(A^c) ]^k} dx_1 \dots dx_k  \tilde j_{i_1\dots i_k} (x_1,\dots,x_k)    {\cal O}^{i_1}(x_1) \dots {\cal O}^{i_k}(x_1)
\, ,
\end{align}
where we have used the fact that operator mixing is suppressed by powers of $1/N$ and defined
\begin{equation}
\mathcal{O}_A = \int_{D(A)} j_i^A(x) {\cal O}^i_A(x), \ \ \ \mathcal{O}_{A^c} = \int_{D({A^c})} j_i^{A^c}(x) {\cal O}^i_{A^c}(x).
\end{equation}
The combinations of operators represented by  ${\cal O}_{A}, {\cal O}_{A^c}$ will play important roles below. Although not explicitly indicated, we will make use of a similar decomposition of the multi-trace parts of \eqref{eq:piAtLAdS} into sums of products of operators associated separately with $A$ and $A^c$. Since \eqref{coherent-state} requires time-ordered products of such operators, we should really perform an expansion of the form \eqref{eq:piAtLAdS} to express the source $J(t)$ at each time in terms of operators that evolve to the regions $A, A^c$, but to avoid clutter in our notation we will not explicitly indicate the time at which given operators are to act.

In terms of a holographic bulk dual, the rewriting of \eqref{coherent-state} as \eqref{eq:piAtLAdS} could be described as evolving sources to a bulk Cauchy surface intersecting the boundary at $A \cup A^c$ and then using e.g. the methods of
\cite{Hamilton:2005ju,Hamilton:2006az,Kabat:2011rz,Heemskerk:2012mn,Kabat:2013wga,Morrison:2014jha,Kabat:2015swa}
to write the corresponding bulk operators in terms of boundary operators\footnote{Since our analysis is perturbative, this construction may be performed in empty AdS so that ball-shaped regions $A, A^c$ define Rindler-AdS regions with a common bifurcate Rindler horizon.} in  $D(A), D(A^c)$. At least for operators ${\cal O}^i$ not associated with bulk gauge fields, this interpretation nicely side-steps issues (see e.g. \cite{Buividovich:2008gq,Donnelly:2011hn,Casini:2013rba,Casini:2014aia,Donnelly:2014gva,Donnelly:2014fua,Hung:2015fla,Ghosh:2015iwa,Aoki:2015bsa,Donnelly:2015hxa}) associated with the expectation that our CFT will be a gauge theory so that its Gauss-law constraint forbids factorization of its Hilbert space into separate Hilbert spaces for $A$ and $A^c$.

However, even without a Gauss-law constraint, quantum field theory Hilbert spaces do not admit a precise tensor product structure.  This fact is associated with singularities in various $n$-point correlation functions which can give contributions localized precisely on the boundary $\partial A$ where $A$ and $A^c$ meet.\footnote{In the bulk description mentioned above, it is associated with the fact that operators at the bifurcation surface of the Rindler horizon cannot be mapped to either the $D(A)$ or $D(A^c)$.}  The issue could be ignored if we were to consider only correlators smeared with smooth functions; we could simply deform the smearing functions so that they vanish in some small neighborhood of $\partial A$ and then recover the original undeformed correlators by continuity as these neighborhoods shrink to zero size.  But the fact that $H_A, H_{A^c}$ are integrals of the stress tensor against non-smooth functions\footnote{The first derivative of the smearing function is discontinuous at $\partial  A$.} means that more care will be required.  In section \ref{sec:computation} below we simply restrict to sources that induce changes in energy $\Delta \langle H_A \rangle$ and entropy $\Delta S_A$ that can be approximated by replacing each term in \eqref{eq:piAtLAdS} with a source that differs from the original in some neighborhood $A_{\t{collar}}$ of $\partial A$ and vanishes smoothly in a smaller neighborhood $A_{\t{source-free}}$ of $\partial A$,  and then letting the width of $A_{\t{collar}}$ vanish.  Roughly speaking, these are the sources that do not produce distributional terms localized at $\partial A$.  We refer to such sources as having non-singular action on the modular Hamiltonian $H_A$.

In practice, rather than working through the above limit explicitly, we will simply consider the regulated operators mentioned above which we take to vanish in a common neighborhood $A_{\t{source-free}}$ of $\partial A$.  The limit $A_{\t{collar}}, A_{\t{source-free}} \rightarrow \emptyset $ will be left implicit. Such sources may for example be obtained by choosing the original stress-tensor sources to be supported in the interior of $D(A) \cup D(A^c)$.  Nevertheless, the fact that the multi-trace terms in either \eqref{coherent-state} or \eqref{eq:piAtLAdS} are multi-local, means that they can include products of operators in $D(A)$ with those in $D(A^c)$.  Thus $U$ is generally not a product of separate unitary transformations on $D(A)$ and $D(A^c)$.

\section{Results with no singular terms} \label{sec:computation}

We now study the changes in energy and entropy associated with applying $U$ to $|0\rangle$ for the above sources.

\subsection{Energy} \label{sec:Energy}

Changes in the energy are straightforward to evaluate and take the form
\begin{align} \label{eq:resultH}
\Delta\langle H_A\rangle  &= \langle \bar{T} e^{i \alpha  J} H_A  T e^{-i \alpha  J} -H_A \rangle \cr
 &= i \epsilon N  \left<[ J, H_A ]  \right>   -  \frac{\epsilon^2 N^2}{2}  \left<[J, [J, H_A]]_{T} \right>  + O(\epsilon^3 N^2),
\end{align}
where $\bar T$ denotes anti-time-ordering and $[J, [J, H_A]]_{T} $ is the operation defined by the second order expansion of the first line; as indicated by the notation, one may think of this term as an appropriately time-ordered version of a double commutator.

Since the first order term also clearly follows from the first line, it remains only to show that the omitted terms are of order $\epsilon^3 N^2$ and in particular involve no higher powers of $N$.  We expect that this argument is also standard, but we state it here for completeness.  The point is to note the repeated commutator structure of $\bar{T}e^{i \alpha  J} H_A  T e^{-i \alpha  J} -H_A$, which requires that at order $\alpha^n$ all $n$ sources must be connected to each other and also to $H_A$.   Considering for the moment only the single-trace contribution ${\cal O}_A + {\cal O}_{A^c}$ to all sources $J$ from \eqref{eq:piAtLAdS} and using \eqref{eq:tHooftRule} then gives only terms of order $\epsilon^n N^2$.

One may then show that multi-trace contributions are further suppressed by at least an additional $N^{-2}$:  Since the coefficient of each multi-trace contribution to \eqref{eq:piAtLAdS} contains an explicit factor of $1/N$, the only possible exception could come from including a single double-trace contribution ${\cal O}^{i_1}{\cal O}^{i_2}$ to one of the sources.  But since $\langle {\cal O}^{i_1} \rangle = \langle {\cal O}^{i_2} \rangle =0$, the repeated commutator structure again means that all that all non-zero contributions are fully connected.  From \eqref{eq:tHooftRule}, the replacement of a single-trace operator by ${\cal O}^{i_1}{\cal O}^{i_2}$ thus yields an extra $1/N$ in this connected correlator giving the stated suppression by two powers of $1/N$.

We should also comment further on the linear term in \eqref{eq:resultH}.  Recall that we take sources in \eqref{eq:piAtLAdS} to vanish near $\partial A$.  This must in particular be true of the single-trace term, which can thus be written as ${\cal O}_A + {\cal O}_{A^c}$.
Writing $H_A$ as an integral of $N {\cal O}^0$ (plus a constant) and using \eqref{eq:tHooftRule}
would suggest that this single-trace part makes the linear term of order $\epsilon N^2$.  But
the ${\cal O}_{A^c}$ term clearly commutes with $H_A$. And since $[{\cal O}_{A} , H_{A^c}] =0$ we have $[{\cal O}_{A} , H_{A}] =[{\cal O}_{A} , K]$.  The fact that $K$ annihilates the vacuum then implies $\left< [{\cal O}_{A} , H_A] \right> =0$.  As a result, the linear term in \eqref{eq:resultH} receives contributions only from multi-trace terms in \eqref{eq:piAtLAdS} and is thus suppressed by an extra $N^{-2}$ as described above.  Thus the linear term is in fact of order $\epsilon N^0$.

\subsection{Entropy}

Computing changes in entropy and relative entropy is more complicated than computing $\Delta \langle H_A \rangle$.  The main issue is that expanding the logarithm in $S_A$ requires using the highly non-trivial Baker-Campbell-Hausdorf (BCH) formula. Nevertheless, subject to the assumption that all sources in \eqref{eq:piAtLAdS} vanish near $\partial A$, we will argue below that changes in the entropy $S_A$ and relative entropy $R_A$ take the form
\begin{subequations} \label{eq:results}
\begin{align}
R_A  &: =
\Delta\langle H_A\rangle - \Delta S_A =
\left< e^{H_A} \Delta \rho_A \lb f(K) - f'(K) \rb  \Delta \rho_A \, e^{H_A} \right> + O(\epsilon^3 N^2)\, , \label{eq:resultR1}\\
&=  \epsilon^2 N^2    \left<   {\cal O}_A {K}  {\cal O}_A \right>  + O(\epsilon^3N^2) + O(\epsilon^2N^0)  \,
\label{eq:resultR2} \\
\Delta S_A  &= O(\epsilon N^0)
\label{eq:resultS}
\end{align}
\end{subequations}
where $\Delta \rho_A = \rho_A - \rho_{A0}$ and $f$ is the smooth function
\begin{align} \label{eq:fdef}
f(y) = \dfrac{y}{1-e^{-y}} \, .
\end{align}
This $f$ appears because it is the generating function of the Bernoulli numbers which play an important role in the BCH formula~\cite{Grensing:1986bg}.  While it is generally redundant to describe $\langle H_A \rangle$, $S_A$, and $R_A : = \ev{H_A} - S_A$, we choose to do so -- and in fact give two expressions for $R_A$ -- both in order to state definite results and to describe the relative sizes of various contributions.  For example, the explicit term in \eqref{eq:resultR1} is a general result for the second order term in the relative entropy $R_A$ of any bipartite quantum system,\footnote{While continuum field theories are not truly bipartite, we may nevertheless use this formula due to our assumption that sources vanish near $\partial A$.   See appendix \ref{app:RA} for details.} but the large $N$ structure is more apparent from \eqref{eq:resultR2}.  Since our current interest focuses on the latter, and since \eqref{eq:resultR2} may be derived from more general arguments,  we relegate the calculation leading to \eqref{eq:resultR1} to appendix \ref{app:RA}.  Equation \eqref{eq:resultR1} may nevertheless be useful for applications that require subleading terms of order $\epsilon^2 N^0$ as well as the leading term of order $\epsilon^2 N^2$.

Although deriving these results will require some work, the intuition behind \eqref{eq:resultR2} and \eqref{eq:resultS} is easy to understand. Note that keeping only the single-trace term would make $U$ a product of separate (commuting) unitary transformations $U^A$ and $U^{A^c}$ on $A$ and $A^c$.  This would require $\Delta S_A =0$ and thus $ R_A   = \Delta\langle H_A\rangle$. So contributions to $\Delta S_A$ require the multi-trace terms in \eqref{eq:piAtLAdS}, which are of order $N^0$.  Deriving \eqref{eq:resultS} thus amounts to controlling cross-terms involving both single-trace and multi-trace source terms.  This is done in section \ref{sec:multi} below.

The explicit term in \eqref{eq:resultR2} is the second-order effect of purely single-trace sources on $R_A$, or equivalently on $\langle H_A \rangle$.   Since either can depend only on the restriction of the state to $A$, they are unchanged by the action of the $U^{A^c}$ defined above.  We may thus consider only $U^A$, which commutes with $H_{A^c}$ and thus induces identical changes in both $H_A$ and $K$.  Indeed, since the linear term in \eqref{eq:resultH} vanishes for single-trace sources that vanish near $\partial A$, it suffices to compute only the second-order term
\begin{equation}
-  \frac{\epsilon^2 N^2}{2}  \left<[\mathcal{O}_A, [\mathcal{O}_A, K]]_{T} \right>.
\end{equation}
Using $K |0\rangle = \langle 0 |K = 0$ then shows the only non-zero term to be the one displayed in \eqref{eq:resultR2}.  Along with the fact that the first order change in $R_A$ vanishes identically by the first law, the errors in \eqref{eq:resultH} and \eqref{eq:resultS} then imply those in \eqref{eq:resultR2}.

\subsection{Multi-trace contributions to $\Delta S_A$}
\label{sec:multi}

The task that remains is to show that including multi-trace contributions to \eqref{eq:piAtLAdS} can change $S_A$ only by terms of order $N^0$. The argument is somewhat lengthy, so we break it into several parts.  We first reorganize the action of the single-trace source-terms in order to show that they have little impact.  We then work to write a series expansion of $S_A$ to which we can usefully apply the large-$N$ counting rule \eqref{eq:tHooftRule}.  Here there are two difficulties, one of which is associated with the fact that $S_A$ is defined as a non-linear function of the reduced density matrix $\rho_A$ on region $A$ and not as a vacuum expectation value.  Any power series expansion of $S_A$ thus naturally involves many traces over $A_c$ that must be eliminated in order to use \eqref{eq:tHooftRule}.  The other involves controlling contributions from possible disconnected correlators.  After completing these tasks, we combine the results and count powers of $N$.

\subsubsection{Reorganizing the action of single-trace sources}
\label{sec:single}

To begin, recall the definitions
\begin{gather} \label{eq:RhoA2}
 \rho_A = \tr_{A^c}  \sigma_\alpha    \,  = \tr_{A^c} \lp Te^{-i\alpha J} \sigma_0 \bar{T} e^{i\alpha J} \rp \;  . \end{gather}
 Recall also that dropping the multi-trace contributions and computing only
\begin{equation}
U_{\text{single}} = T e^{-i \alpha \int dt J_{\text{single}}}
\end{equation}
would give a product of separate unitary transformations on $A$ and $A^c$  that do not change $S_A$.  Similarly, we may note that the entropy of $\rho_A$ must be identical to the entropy of
\begin{gather} \label{eq:RhoAc}
 \rho^{\t{conj}}_A = \tr_{A^c} \lp  U^{-1}_{\text{single}}  \sigma_\alpha U_{\text{single}}  \rp
 =  [U^{A}_{\text{single}} ]^{-1} \rho_A U^A_{\text{single}},
 \end{gather}
where $ U^A_{\text{single}}$ is the part of  $U_{\text{single}}$ that acts on $A$.  We thus have
\begin{gather} \label{eq:SA2}
S_A = - \tr_{A} \left( \rho^{\text{conj}}_A  \log \rho^{\text{conj}}_A  \right) .
\end{gather}

Now, $\sigma_\alpha$ itself is defined (see \eqref{coherent-state}) by conjugating $|0\rangle \langle 0|$ with a time-ordered unitary $U$.  This unitary is the product of many factors of the form
\begin{equation}
\label{factor}
e^{-i \alpha J(t) \delta t} =
 e^{-i \alpha J_{\text{single}}(t) \delta t} e^{-i \alpha J_{\text{multi}}(t) \delta t} +O(\delta t^2)
 \end{equation}
where we have separated the single- and multi-trace parts of the source at time $t$.
Note that our new conjugation  by $U_{\text{single}}$ merely replaces the
$U$ in \eqref{coherent-state}  by $U_{\text{single}}^{-1} U$, or equivalently replaces each factor \eqref{factor} with
\begin{equation}
\label{factor2}
e^{-i \alpha J^{\text{conj}}_{\text{multi}}(t) \delta t},
 \end{equation}
where $J^\text{conj}_\text{multi}(t)$ is a multi-trace term conjugated by a time-ordered exponential built from single-trace sources at earlier times; there is no effect on \eqref{factor} from sources at later times as these merely cancel between $U^{-1}_{\text{single}}$ and $U$. One may make an analogy between $U$ and the time-ordered exponential that implements Heisenberg-picture time-evolution in some quantum system, with $J_{\text{single}}$ playing the role of the free Hamiltonian and $J_{\text{multi}}$ playing the role of the interaction terms.  Our conjugation by $U_{\t{single}}$ then plays the role of passing to the interaction picture, where the new time evolution is a product of factors like \eqref{factor2}.

We wish to write
\begin{equation} \label{eq:Drform}
\rho^{\text{conj}}_A = \rho_{A0} \lp 1 + \rho_{A0}^{-1} \Delta \rho^{\text{conj}}_A \rp
\end{equation}
for
\begin{gather} \label{eq:DeltaRhoAc}
\Delta \rho^{\text{conj}}_A =  \rho^{\text{conj}}_A -  \rho_{A0} =
\tr_{A^c} \lp  U^{-1}_{\text{single}}  \sigma_\alpha U_{\text{single}} - \sigma_0 \rp,
\end{gather}
and to expand \eqref{eq:SA2} in powers of $\rho_{A0}^{-1} \Delta \rho^{\text{conj}}_A$.  We then expand $\Delta \rho^{\text{conj}}_A $, in powers of $\alpha$. In this latter step will write each $\Delta \rho_A^\t{conj}$ as a sum of terms of the form
\begin{equation}
\label{eq:conj}
\tr_{A^c} \lp [J^\text{conj}_\text{multi-1}, [J^\text{conj}_\text{multi-2} \dots [ J^\text{conj}_\text{multi-$k$}, |0 \rangle \langle 0| ]] \dots ] \rp,
\end{equation}
where $J^\text{conj}_\text{multi-$j$}$ is a time integral of  $J^\text{conj}_\text{multi}(t)$ and the factors are appropriately time-ordered.

It is important to note that $|0 \rangle \langle 0|$ is not conjugated by any such unitary; the privileged position of this operator at the end of the chain of repeated commutators means that, in our analogy with transforming between the Heisenberg and interaction pictures, this operator can be thought of as labeled by the earliest possible time so that the picture-changing transformation acts on it trivially.  The point of the form \eqref{eq:conj} is that we will shortly (see section \ref{sec:useful}) transform the expansion of \eqref{eq:SA2} into a form involving only vacuum correlators of products of operators. The repeated commutator structure of $J^\text{conj}_\text{multi-$j$}$ will then forbid single-trace terms from appearing in correlators unless they are appropriately connected to multi-trace terms.

\subsubsection{The entropy as a correlator}
\label{sec:useful}

The next step in our argument is to show how $S_A$ can be written as a sum of vacuum correlators of products of the ${\cal O}^i$ with functions of the boost Hamiltonian $K$, and where the remaining $N$-dependence of each term follows directly from powers of $\alpha = \epsilon N$ and \eqref{eq:piAtLAdS}. In particular, all extra traces over $A^c$ will be removed, and the modular Hamiltonians $H_A, H_{A^c}$ will not appear except in the combination $K = H_A - H_{A^c}$.  This last feature will be critical in controlling contributions from disconnected correlators.

 As a first step toward this goal, we may use \eqref{eq:Drform} to express $S_A$ as the vacuum correlator
\begin{align}
\label{eq:SA6}
S_A &= - \tr_{A} \left[ \rho_{A0} \lp 1 + \rho_{A0}^{-1} \Delta \rho^{\text{conj}}_A \rp \log  \rho^{\text{conj}}_A    \right] \cr
&= - \left\langle  \lp 1 + \rho_{A0}^{-1} \Delta \rho^{\text{conj}}_A \rp \log \rho^{\text{conj}}_A \right\rangle,
\end{align}
where operators on region $A$ are to be interpreted as operators on the full CFT Hilbert space by tensoring them with the identity on $A^c$.  Note that $\log \lp \rho^{\text{conj}}_A \otimes \mathds{1}_{A^c} \rp = \lp \log \rho^{\text{conj}}_A \rp + \log \mathds{1}_{A^c} = \lp \log \rho^{\text{conj}}_A \rp$, so that we may interpret $ \log \rho^{\text{conj}}_A$ as the logarithm of an operator on the full Hilbert space.

We next remove the explicit traces over $A^c$.  These enter through the definition \eqref{eq:rhodef}, and are potentially problematic because the sources involve products of operators on $A$ with operators on $A^c$.  The trick to proceeding is to use the assumption that each term in \eqref{eq:piAtLAdS} is supported away from $\partial A$ (so that contributions to each source-term from $A$ and $A^c$ commute with each other) to write each term \eqref{eq:conj} as a sum of terms in which all operators on $A^c$ have been commuted to act directly either on $|0\rangle$ from the left or on $\langle 0|$ from the right.

Using the entanglement properties of $|0\rangle$, we may now  replace each operator in $A^c$ by a so-called ``mirror operator'' on $A$; see e.g. \cite{Haag:1992hx}, though the particular terminology is from the more recent \cite{Papadodimas:2013wnh}.  To explain how this works, let us for the moment take $A$ to be the ``southern'' hemisphere of our $S^{d-1}$ at $t=0$. In any relativistic theory, given an  operator ${\cal O}$, not necessarily a scalar or even local, we may study the CPT conjugate operator ${\cal O}^{CPT}$.  Since the parity operation exchanges the north and south hemispheres, the CPT conjugate of any northern hemisphere operator at $t=0$ (i.e., on $A^c$) is an operator on the $t=0$ southern hemisphere $A$.  Furthermore, these operators satisfy
\begin{equation}
\label{eq:CPT}
{\cal O}^{CPT} |0 \rangle = e^{K/2} {\cal O} |0 \rangle; \ \ \ \langle 0 | {\cal O}^{CPT\dagger} = \langle 0|  {\cal O}^\dagger e^{K/2},
\end{equation}
where the second relation is just the Hermitian conjugate of the first.
These are just Kubo-Martin-Schwinger (KMS) relations in terms of the imaginary time evolution generated by our $K$, so they encode the thermal nature of $|0\rangle$ with respect to $H_A$.  Thus the mirror operator $\tilde {\cal O} := e^{-K/2} {\cal O}^{CPT} e^{+K/2} $ satisfies
\begin{equation}
\tilde {\cal O} |0 \rangle =  {\cal O} |0 \rangle; \ \ \ \langle 0 | \tilde {\cal O}^\dagger = \langle 0|  {\cal O}^\dagger e^{K}.
\end{equation}

The key point is that the CPT conjugate of an ${\cal O}^i$ is just another ${\cal O}^j$ (or perhaps a linear combination thereof).   So at the expense of introducing additional factors of $e^{-K/2} = e^{-H_A/2} e^{H_{A^c}/2}$, we may replace the $A^c$ operators acting on $\sigma_0$ by CPT-conjugate operators acting on $A$.   Conformal invariance then guarantees that we can again perform a corresponding operation to replace operators on $A^c$ with those on $A$ for more general ball-shaped regions. The new operators $e^{H_{A^c}/2}$ may then be commuted past $A$-operators to act on $\sigma_0$, where they may be replaced by factors of $e^{H_{A}/2}$ using \eqref{eq:H0}.  The net result is thus to write $\Delta \rho_A$ as a sum of terms of the form
\begin{equation}
\label{eq:term}
\tr_{A^c} \lp {\cal O}_A^{i_1}\dots {\cal O}_A^{i_n} e^{-H_A/2} {\cal O}_A^{j_1}\dots {\cal O}_A^{j_m}      e^{+H_A/2} |0 \rangle \langle 0 |  e^{+H_A/2}
{\cal O}_A^{k_1}\dots {\cal O}_A^{k_r} e^{-H_A/2} {\cal O}_A^{l_1}\dots {\cal O}_A^{l_s}  \rp.
\end{equation}

Operators supported away from $A^c$ can now be pulled outside the trace over $A^c$.  By assumption these include all ${\cal O}_A^i$, but we should take care with the factors of $e^{H_A/2}$ which include support near $\partial A$.  We may do so in each term by using the Zassenhaus formula
\begin{equation}
\label{eq:Z}
e^{X+Y} = e^X e^Y e^{-[X,Y]/2} e^{\frac{1}{6}\lp 2[Y,[X,Y]] +[X,[X,Y]] \rp}\dots
\end{equation}
for $H_A/2 = X + Y$, with $X = H_A^\text{far}/2$ a Hermitian integral of the subtracted stress tensor $N {\cal O}_A^0$ weighted by a smooth function supported away from $\partial A$ and $Y$ a similar integral supported close enough to $\partial A$ to avoid overlap with the support of any ${\cal O}_A^i$ appearing in the given term.  Since both are smooth integrals, the supports of $X$ and $Y$ will overlap with each other and thus yield non-zero commutators in \eqref{eq:Z}.  It is convenient to collect all factors involving $Y$ into a single operator $e^{H_A^\text{near}/2}$.  Note that commutators of $X,Y$ receive contributions only from regions in the support of both operators, so that $e^{H_A^\text{near}/2}$  is again supported in a region close to $\partial A$ that avoids overlap with the support of any ${\cal O}_A^i$ appearing in the given term.

We now have
\begin{subequations}
\label{eq:far}
\begin{align}
e^{-H_A/2} &= e^{-H_A^\text{far}/2} e^{-H_A^\text{near}/2}  ,
\label{eq:Z-}\\
e^{+H_A/2} &= e^{+H_A^\text{near}/2} e^{+H_A^\text{far}/2}    \label{eq:Z+} \\
e^{-H_A/2} &= e^{-H_A^{\text{near}\dagger}/2} e^{-H_A^\text{far}/2}, \label{eq:Z-dag}\\
e^{+H_A/2} &=  e^{+H_A^\text{far}/2} e^{+H_A^{\text{near}\dagger}/2} ,
\label{eq:Z+dag}
\end{align}
\end{subequations}
where \eqref{eq:Z+} is the inverse of \eqref{eq:Z-} and the final two relations \eqref{eq:Z-dag}, \eqref{eq:Z+dag} are the adjoints of \eqref{eq:Z-}, \eqref{eq:Z+}.  The expressions \eqref{eq:far} allow us to safely reformulate \eqref{eq:term} as
\begin{align}
\label{eq:term2}
&{\cal O}_A^{i_1}\dots {\cal O}_A^{i_n} e^{-H_A^\text{far}/2} {\cal O}_A^{j_1}\dots {\cal O}_A^{j_m}
\cr & \qquad \times
\tr_{A^c} \lp e^{-H^\text{near}_A/2}  e^{+H_A^\text{near}/2}  e^{+H_A^\text{far}/2} |0 \rangle \langle 0 | e^{+H_A^\text{far}/2} e^{+H_A^\text{near}/2}    e^{-H_A^\text{near}/2}\rp \cr
& \qquad \qquad \times {\cal O}_A^{k_1}\dots {\cal O}_A^{k_r} e^{-H_A^\text{far}/2} {\cal O}_A^{l_1}\dots {\cal O}_A^{l_s}  .
\end{align}
Since $e^{-H_A^\text{near}/2}$ is the inverse of $e^{H_A^\text{near}/2}$, these operators all cancel.  The remaining factors of $e^{+H_A^\text{far}/2}$ may then be extracted from the trace as well.  Reversing the steps involving $H_A^\text{near}, H_A^\text{far}$ then reconstructs the original factors of $e^{\pm H_A/2}$ outside the trace, leaving just \eqref{eq:term} where the trace acts only on $|0\rangle \langle 0 |$ as one would naively expect.  Using $\tr_{A^c}  |0\rangle \langle 0|  = e^{-H_A}$ then removes all explicit traces over $A^c$.  While we could take this final factor of $e^{-H_A}$ to cancel the two factors of $e^{H_A/2}$ adjacent to $|0\rangle\langle0|$ in \eqref{eq:term}, due to the structure of \eqref{eq:conj} it is convenient to instead keep all of these factors explicit.  The latest factor of $e^{-H_A}$ coming from the trace is distinguished, and we shall refer to it below as the `final' $e^{-H_A}$.

The above steps have written each factor of $\Delta \rho^{\text{conj}}_A$ as a product of ${\cal O}^i_A$ with two factors of $e^{-H_A/2}$, two factors of $e^{+H_A/2}$, and our final $e^{-H_A}$.  But as noted above we wish to expand in powers of $\rho_{A0}^{-1} \Delta \rho^{\text{conj}}_A = e^{H_A} \Delta \rho^{\text{conj}}_A$.  We may then use
\begin{equation}
\mathds{1} = e^{H_{A^c}} e^{-H_{A^c}} e^{-H_{A^c}/2} e^{-H_{A^c}/2}  e^{H_{A^c}/2} e^{H_{A^c}/2}
\end{equation}
to insert the indicated factors of  $e^{H_{A^c}/2}$. Commuting them past the ${\cal O}^i_A$ and exponentials of $H_A$ as needed allows us to replace all exponentials of $H_A$ with exponentials of $K = H_A - H_{A^c}$.  In particular, in this way we obtain a distinguished `final' factor of $e^{-K}$.

We are now close to achieving our goal.  If $\rho^{-1}_{A0} \Delta \rho^{\text{conj}}_A$ would commute with $\rho_{A0}$, we could use
$\log \rho_A^{\text{conj}}  - \log \rho_{A0}= \log \lp 1 + \rho_{A0}^{-1} \Delta \rho_A^{\text{conj}} \rp$
and expand \eqref{eq:SA6} in a standard Taylor series.  Combined with our results above, this would give an infinite sum of terms with each being a vacuum correlator of a product of ${\cal O}_i$ and exponentials of $K$, consistent with the form required above.\footnote{With the exception of the term $\langle \rho^{-1}_{A0} \Delta \rho^{\text{conj}}_A H_A \rangle$.  This term resembles the mulit-trace contributions to $\Delta\langle \log \rho_{A0}\rangle$ considered in section~\ref{sec:Energy} and is also of order $N^0$ for similar reasons.}

But since $\rho^{-1}_{A0} \Delta \rho^{\text{conj}}_A$ and $\rho_{A0}$ generally do not commute we must use the Baker--Campbell--Hausdorff formula to evaluate $\log \rho^{\text{conj}}_A = \log \left[ \rho_{A0} \left(1 + \rho_{A0}^{-1} \Delta \rho^{\text{conj}}_A \right) \right]$.  An explicit formulation of this identity due to Dynkin takes the form (see e.g.~\cite{rossmann2006lie})
\begin{align}
\label{Dynkin}
\log\lp e^Xe^Y\rp = \sum_{k=1}^\infty \frac{(-1)^{k-1}}{k} \sum_{p_i+q_i\geq 1} \frac{\lb X^{(p_1)} Y^{(q_1)}\dots X^{(p_k)}Y^{(q_k)}\rb}{ \lp \sum_{i=1}^k (p_i+q_i) \rp  \lp \prod_{i=1}^k p_i! q_i ! \rp } \, ,
\end{align}
where
\begin{align}
\lb X^{(p_1)} Y^{(q_1)}\dots X^{(p_k)}Y^{(q_k)}\rb \equiv [\underbrace{X,[X,\dots[X}_{\t{$p_1$ $X$s}},[\underbrace{Y,\dots[Y}_{\t{$q_1$ $Y$s}},[\underbrace{X,\dots[X}_{\t{$p_k$ $X$s}},[\underbrace{Y,[Y,\dots Y}_{\t{$q_k$ $Y$s}}]\dots] \, .
\end{align}
We wish to set $X= -H_A = \log \rho_{A0}$ and $Y = \log \left(1 + \rho_{A0}^{-1} \Delta\rho^{\text{conj}}_A \right)$.

The important observation is that the form of $\rho_{A0}^{-1} \Delta \rho_A $ found above clearly commutes with $H_{A^c}$, so  $[H_{A^c},Y]=0$ as well.  Recalling from \eqref{commute} that $X=-H_A$ also commutes with $H_{A^c}$ allows us to replace each $X$ in the above repeated commutators with $-K = H_{A^c} - H_A$.  The result is an expression for $S_A$ as an infinite sum over terms, each of which is a vacuum correlator involving only products of ${\cal O}^i_A$ and functions of $K$, and with all further explicit dependence on $N$ coming from \eqref{eq:piAtLAdS} as desired.

\subsubsection{Counting powers of $N$}

We are now ready to count powers of $N$.  Consider then contributions at order $\alpha^{p}$ involving $r$ single-trace operators ${\cal O}_\text{from \ single}$ that come from single-trace source-terms and $s$ single-trace operators ${\cal O}_\text{from \ multi}$ that come from multi-trace source-terms.  Such a contribution is equal to a product of connected correlators each with $r_i$ of the ${\cal O}_\text{from \ single}$ and $s_i$ of the ${\cal O}_\text{from \ multi}$ such that $r = \sum_i r_i$ and $s = \sum_i s_i$.  We need not keep track of the factors of $K$ since each contributes an explicit factor of $N$ that is cancelled by the extra $1/N$ from \eqref{eq:tHooftRule} associated with the ${\cal O}^0$ it contributes to any correlator.  Each connected correlator gives $r_i +s_i -2$ factors of $1/N$ by~\eqref{eq:tHooftRule}. The single-trace sources provide an additional $r$ explicit powers of $N$.  For double-trace sources the explicit $N$ from $\alpha = \epsilon N$ cancels against the $1/N$ in \eqref{eq:piAtLAdS} and for higher-trace sources the contribution is more suppressed.  So the total number of factors of $1/N$ is greater than or equal to $\sum_i (s_i - 2)$.  We wish to show that this sum is non-negative.

Recall that the reason multi-trace sources affect the entropy is that they can contain products of operators in $A$ with operators in $A^c$.  But a given term will only be sensitive to the correlations created by these sources if at least one operator from $A$ and one from $A^c$ appear in the same connected correlator.  So it is natural to expect that the entropy will only receives contributions from terms with $s_i\ge2$ for all $i$, which would imply the desired result.

To show that this is so, we note that there are no contributions from terms with $s_i = 0$.  This is because all of the ${\cal O}_\text{from \ single}$ occur in nested commutators inside some $J^\text{conj}_\text{multi-$j$}$ and therefore must be connected to at least one ${\cal O}_\text{from \ multi}$ in order to contribute.

Now suppose $s_i = 1$, where we take the relevant ${\cal O}_\text{from \ multi-$j$}$ to come from a $m$-trace source term living inside some $J^\text{conj}_\text{multi-$j$}$. Since $\langle {\cal O}_\text{from \ multi-$j$} \rangle =0$, a non-vanishing connected correlator must involve other operators. None of these can come from single-trace source terms in other  $J^\text{conj}_\text{multi-$k$}$ for $k\neq j$, which must instead stay attached to their own multi-trace sources.  So since $s_i =1$, if any of these operators come from a final $K$ then the correlator must vanish due to $K|0 \rangle =0 = \langle 0 | K$ and the fact that \eqref{eq:conj} prohibits a final $K$ from intervening between factors coming from any given $J^\text{conj}_\text{multi-$j$}$.  So the correlator consists only of ${\cal O}_\text{from \ multi-$j$}$, single-trace source-terms, and $K$'s coming from a single $J^\text{conj}_\text{multi-$j$}$.  Terms of this form exponentiate, and have the effect of replacing the operator ${\cal O}_\text{from \ multi-$j$}$ in $J^\text{conj}_\text{multi-$j$}$ with a classical source of order $N$, demoting $J^\text{conj}_\text{multi-$j$}$ to an $(m-1)$-trace source of order $N^{-{m-3}}$ (i.e., the same order as the $(m-1)$-trace sources already appearing in \eqref{eq:piAtLAdS}).  So we need not consider such terms separately when counting powers of $N$.  In particular, double-trace sources demoted to single-trace sources in this way
can be absorbed into the $U_{\t{single}}$ of section \ref{sec:single} and do not contribute to the entropy.  So all relevant terms have $s_i \ge 2$ and the largest possible contribution to the entropy is of order $N^0$ as anticipated above.

\section{Discussion} \label{sec:discussion}

The work above contains first steps toward studying the von Neumann entropy of excited states in CFTs satisfying the 't Hooft large $N$ counting rule \eqref{eq:tHooftRule}.  We considered the entropy of ball-shaped regions for states produced by real-time sources with non-singular action on the modular Hamiltonian and of the form \eqref{eq:piAtLAdS}.  From a dual bulk point of view, such sources produce small classical waves -- or more properly quantum coherent states with large amplitude of order $\epsilon N$ -- on both sides of a Rindler horizon but which do not disturb the bifurcation surface itself.  They also add $O(N^0)$ entangled sets of particles across the horizon.  As the former does not change the HRT entropy, one expects that the CFT entanglement changes only at order $N^0$.  We have verified that this is indeed the case by a direct argument in the CFT.  While our results are directly formulated in terms of a $1/N$ expansion, it may be interesting to follow \cite{Faulkner:2014jva} and attempt to formulate a version of our results that would hold in arbitrary CFTs, whether or not they have a large-$N$ counting rule like \eqref{eq:tHooftRule}.\footnote{We thank Tom Faulkner for discussion on this point.}

One consequence of our work is the explicit formula \eqref{eq:resultR2} for the order $\epsilon^2 N^2$ relative entropy.   When interpreted in the bulk theory, this formula is precisely the bulk stress tensor on the $A$ side of the bulk horizon, integrated so as to give the associated change in the boost energy $K$. This may be seen from the fact that \eqref{eq:resultR2} is just the contribution to $K$ from the excitations on the $A$ side of the bifurcation surface, and by bulk causality that it thus gives the value of the boost Hamiltonian $K$ in a related bulk solution produced from the vacuum using only the sources ${\cal O}_A$ in the causal past of $A$.  The above claim then follows by writing $K$ in terms of a standard `bulk stress tensor' that includes quadratic contributions from gravitons as well.  For concreteness, one may choose the graviton contribution to be given by the AdS analogue of the Landau-Lifshitz `pseudo-tensor' expression \cite{Landau:1987gn}, though our requirement that perturbations vanish near the bifurcation surface means that many other choices give equivalent results\footnote{There is, however, a preferred expression when perturbations do not vanish at the bifurcation surface. The bulk relative entropy $H_A - S$ of the classical bulk is then the so-called canonical energy density of \cite{Hollands:2012sf}, which is naturally expressed as an integral of what one may call a bulk stress tensor.  See ~\cite{Lashkari:2015hha} for further discussion of this connection.}.

The above observation means that our analysis provides a new argument for the universal coupling of gravity to all classical forms of stress-energy.  Though it differs in detail from \cite{Swingle:2014uza}, this interpretation of our classical $(O(N^2))$ result is inspired by the quantum $(O(N^0))$ argument of that reference and reinforces the connection found there between the universal structure of variations in $S_A$ and the universal coupling of bulk gravity.

Another consequence is to give a derivation of the order $N^0$ Faulkner-Lewkowycz-Maldacena correction \cite{Faulkner:2013ana} to the Ryu-Takayagi and Hubeny-Rangamani-Takayanagi conjectures (for ball-shaped regions $A$). In the semi-classical bulk, this argument is perturbative in departures from empty AdS.  It thus complements the original reasoning in \cite{Faulkner:2013ana} in that it does not rely on the Lewkowycz-Maldacena argument~\cite{Lewkowycz:2013nqa}, or on any other use of the replica trick.  At first order in $\epsilon$ the result follows from the first law, but the arguments of section \ref{sec:computation} also give this result at higher orders.   In general, we may decompose $S_A$ into a first-law-piece and $R_A$.  Recall that at second order $R_A$ is quadratic in the first order contribution to $\Delta \rho_A$ and in particular is given by
\begin{equation}
-i\alpha\tr_{A^c} \lp [J_\text{double}, |0\rangle \langle 0| ] \rp
= \tr_{A^c} \lp e^{-i\alpha J_\text{double}} |0\rangle \langle 0| e^{i\alpha J_\text{double}} - \ket{0}\bra{0} \rp + O(\epsilon^2),
\end{equation}
where $J_{\text{double}}$ is the double-trace part of \eqref{eq:piAtLAdS}.  As before the only terms in $J_{\text{double}}$ that can contribute to $\Delta S_A$ are those that involve one operator in $A$ and one in $A^c$.  Such terms throw entangled pairs into the bulk, with one member of each pair on each side of the bulk horizon.  Given the agreement of bulk and CFT vacuum correlators, it is manifest that the corresponding change in bulk entanglement at this order can be computed just as we have done in the CFT and that the results agree.  A similar argument holds at higher orders in $\epsilon$, where the repeated commutator structure of \eqref{eq:conj} reproduces the effect of propagating small effects from the multi-trace sources through the large semi-classical coherent state produced by the single-trace parts of \eqref{eq:piAtLAdS}.  It also extends to higher orders in $N$ to argue that bulk entanglement gives the full series of $1/N$ corrections to HRT.  Note, however, that since our sources are confined to $D(A) \cup D(A^c)$ all perturbations vanish at the bulk HRT surface. Our derivation is thus insensitive to possible perturbative shifts of the HRT surface of the kind predicted in~\cite{Engelhardt:2014gca}.  Conversely, extending our results to allow sources supported on $\partial A$ would in principle allow us to test the conjecture of~\cite{Engelhardt:2014gca} that the full CFT entropy is given by the generalized entropy of a quantum extremal surface, defined as the bulk surface that extermizes the bulk generalized entropy.

We have focussed on ball-shaped regions for simplicity, but we expect our arguments to generalize to arbitrary regions $A$.  Indeed, as described in \cite{Rosenhaus:2014woa}, one may address perturbative deformations of ball-shaped regions by inserting additional factors of the stress tensor (or, equivalently, of $N{\cal O}^0$).  These insertions tend to add another operator to each connected correlator, giving an extra $1/N$ from \eqref{eq:tHooftRule} that cancels the explicit new factor of $N$.  A similar argument can be used to compute corrections to \eqref{eq:CPT} and construct the relevant ``mirror operators'' for these deformed regions.  So the only ingredients of our analysis that remain to be checked are that $H_A$, $H_{A^c}$ can be thought of as operators on the full CFT Hilbert space that commute both with each other, and with local operators supported away from $\partial A$.  It should be possible to analyze these assumptions perturbatively as well.  We expect that these will indeed hold at this level, but verifying this is beyond the scope of our work.

The extension to include sources supported near $\partial A$ would clearly be of great interest.  At linear order in $\epsilon$ the contribution to $S_A$ from such sources is governed by the first law and was effectively studied in \cite{Lashkari:2013koa,Faulkner:2013ica}.  But at this order there is no displacement of the bulk extremal surface.  In particular, for ball-shaped regions $A$ the bulk HRT surface continues to coincide with the bifurcation surface of the corresponding event horizon; i.e., with the causal holographic information surface of \cite{Hubeny:2012wa}, which is expected to compute some coarse-grained version of the CFT entropy \cite{Hubeny:2012wa,Freivogel:2013zta,Kelly:2013aja}.  In contrast, the two are distinguished at second order, so comparing bulk and CFT computations may give insight into the nature of the relevant coarse-graining.  In particular, one might hope to either support or falsify the conjecture \cite{Kelly:2013aja} that it corresponds to the maximizing the entropy over all states for which certain one-point functions coincide with the original state.

In addition, comparing bulk and CFT computations to second order would derive or falsify the HRT conjecture at a non-trivial level.  Even for static perturbations the fact that it avoids the replica trick would make this a useful complement to the Lewkowycz-Maldacena argument~\cite{Lewkowycz:2013nqa}, and the perturbative method should be able to address general time-dependence to which~\cite{Lewkowycz:2013nqa} does not apply.  And since second-order results are sensitive to the displacement of the HRT surface, they could in particular detect any possible motion of this surface in imaginary directions within the complexified AdS spacetime; i.e., they could help diagnose the possible role of complex extremal surfaces as explored in \cite{Fischetti:2014zja}.

One reason that we have avoided such singular terms here is that they are in principle sensitive to the particular way that entanglement is to be defined in the CFT.  Since the CFT is a gauge theory, this can involve a number of subtle issues \cite{Buividovich:2008gq,Donnelly:2011hn,Casini:2013rba,Casini:2014aia,Donnelly:2014gva,Donnelly:2014fua,Hung:2015fla,Ghosh:2015iwa,Aoki:2015bsa,Donnelly:2015hxa}).  Lewkowycz-Maldacena~\cite{Lewkowycz:2013nqa} suggests that the correct notion of CFT entropy is defined by the replica trick, which is precisely the computational tool we wish to avoid.  However, the results of \cite{Buividovich:2008gq,Donnelly:2011hn,Casini:2013rba,Casini:2014aia,Donnelly:2014gva,Donnelly:2014fua,Hung:2015fla,Ghosh:2015iwa,Aoki:2015bsa,Donnelly:2015hxa})  also suggest that the various definitions of entropy differ only by a local boundary term that will cancel in computing the mutual information between pairs of regions $A,B$.  One should thus be able to ignore such concerns in this context. We hope to compute the `singular' second order terms and to explore the above issues in the near future, perhaps using a suitably-generalized version of the calculation in appendix \ref{app:RA}.

\acknowledgments

It is a pleasure to thank David Berenstein, William Donnelly, Eric Dzienkowski, Monica Guica, Tom Hartman, Veronika Hubeny, Ted Jacobson, Nima Lashkari, Juan Maldacena, Mukund Rangamani, Vladimir Rosenhaus, Mark Van Raamsdonk, and Aron Wall for helpful discussions and feedback.  We especially thank Tom Faulkner for his comments on an early draft of this paper. This work was supported by the National Science Foundation under grant numbers PHY12-05500 and PHY15-04541, and by funds from the University of California. In addition, K.K.\ is supported by the NSF GRFP under Grant No.~DGE-1144085. D.M. thanks the Aspen Center and its NSF Grant \#1066293 for their hospitality during the discussions where certain aspects of this project were conceived.  He also thanks the KITP for their hospitality during the final stages of the project, where his work was further supported in part by National Science foundation grant number PHY11-25915.

\appendix

\section{Computing the relative entropy} \label{app:RA}

This appendix derives an explicit formula for the second order change $\delta^2 R_A$ in the relative entropy $R_A$ corresponding to an arbitrary change $\delta \rho_A$ in the reduced density matrix for $A$. This result is not directly used in the main text, other than writing \eqref{eq:resultR1}. Our final expression~\eqref{eq:delta2RA} bears a striking resemblance to Eq.~(C7) of~\cite{Faulkner:2014jva}.  In fact,~\eqref{eq:delta2RA} can also be derived from a straightforward generalization of the calculation leading to (C7).  We present a different, somewhat more involved calculation here because we hope that this approach will be useful for analyzing the additional terms at order $\epsilon^2$ that arise when the sources in \eqref{eq:piAtLAdS} do not vanish in a neighborhood of $\partial A$.

The setting for the calculation below is an arbitrary bipartite quantum system, meaning that it is the tensor product of a system on $A$ and one on $A^c$.  As discussed in the main text, the actual system we study is not strictly of this form, though it can be treated as such at least under our assumption that each term in \eqref{eq:piAtLAdS} vanishes in a neighborhood of $\partial A$.  So a critical step in realizing the hope expressed in the paragraph above is understanding modifications that arise when this assumption fails.

\subsection{The Baker--Campbell--Hausdorff Formula}

We will compute $\delta^2 R_A$ using the BCH formula in the form \eqref{Dynkin}. Since we work only to order $Y^2$, it will be useful to rewrite~\eqref{Dynkin} as
\begin{align} \label{eq:BCH}
\log\lp e^Xe^Y\rp = X + Y + \sum_{n=1}^\infty C_n [X^{(n)}Y^{(1)}] + \sum_{n=1}^\infty \sum_{k=0}^{n-1} D_{n,k} [X^{(k)}Y^{(1)}X^{(n-k)}Y^{(1)}] + O(Y^3) \, ,
\end{align}
where $C_n$ and $D_{n,k}$ are rational numbers.  It was shown in~\cite{Grensing:1986bg} that
\begin{align}
C_n = \frac{(-1)^n B_n}{n!}\, ,
\end{align}
where $B_n$ are the first Bernoulli numbers $(B_0 = 1,B_1= -1/2,\dots )$.  The Bernoulli numbers can be defined by the smooth generating function
\begin{align} \label{eq:generatingFunction}
f(y) := \frac{y}{1-e^{-y}} = \sum_{n=0}^\infty  \frac{B_n}{n!} (-y)^n \, ,
\end{align}
where the sum on the right hand side converges for $y\in(-2\pi,2\pi)$.  Note that $B_{2m+1} = 0$ for $m \ge 1$.

To compute $\delta^2 R_A$ we will also need the coefficients $D_{n,0}$.  To our knowledge these coefficients have not previously been explicitly computed.  But a straightforward application of the recursive technique developed in~\cite{Grensing:1986bg} yields\footnote{Iterating Eq.~(12) of~\cite{Grensing:1986bg} in such a way as to obtain all terms with two $Y$'s gives
\begin{align}
\log(e^X e^Y) = X+Y + \sum_{k=1}^\infty R_k \int_0^1 dt \lp \text{Ad}_X + t \text{Ad}_Y + \sum_{m=1}^\infty R_m [{\text{Ad}_X}^{(m)} t \text{Ad}_Y] + \dots \rp^n Y
\end{align}
where $R_n = (-1)^n B_n/n!$ and $\text{Ad}_Z W = [Z,W]$.  After collecting all of the the terms that give nested commutators of the form $[Y X^{(n)} Y]$ we obtain the first equality in~\eqref{eq:Dn0}.}
\begin{align}  \label{eq:Dn0}
D_{n,0} = -\frac{1}{2(n+1)!} \sum_{k=1}^{n+1} (-1)^{n-k} {n+1 \choose k} B_k B_{n+1-k} = - \frac{B_{n+1}}{2 (n!)} \, .
\end{align}
To obtain the second equality first note that for even $n$ the only non-vanishing terms in the sum are $k=1$ and $k=n$, and that these terms cancel.  For $n=1$ the equality is easily checked by hand.  For odd $n \ge 3$, only the even $k$ terms survive so $(-1)^{n-k} = -1$.  The remaining sum is easily evaluated using a well known identity due to Euler and independently rediscovered by Ramanujan (see for example Eq.~(1.2) in~\cite{Dilcher199623})
\begin{align}
\sum_{j=1}^{m} {2m \choose 2j} B_{2j} B_{2m-2j} = -2m \, B_{2m} \, , \qquad m\ge 2 \, .
\end{align}

\subsection{Computing $\delta^2 R_A$}

Taking a second variation of $R_A$ gives
\begin{align} \label{eq:delta2RAstart}
\delta^2 R_A = \tr \lp \delta\rho_A \, \delta \lb \log(\rho_A) \rb \rp + \tr \lp \rho_A \, \delta^2 \lb \log(\rho_A) \rb \rp \, .
\end{align}
To evaluate the variation of the logarithm we first rewrite
\begin{align}
\log(\rho_A + \delta\rho_A)  = \log \lb e^{-H_A} (1 + e^{H_A} \delta\rho_A)\rb  \,
\end{align}
and apply~\eqref{eq:BCH} with $X=-H_A$ and $Y = \log(1+e^{H_A}\delta\rho_A)$.  We then Taylor expand $Y$ using
\begin{align}
\log\lp 1+ Z  \rp = Z - \frac{Z^2}{2} + O(Z^3)  \, .
\end{align}
None of the $D_{n,k \ge 1 }$ terms from~\eqref{eq:BCH} contribute to~\eqref{eq:delta2RAstart} because they can only appear in the second term of~\eqref{eq:delta2RAstart} and
\begin{align}
&\tr\lp \rho_A \lb H_A^{(k\ge 1)} (\rho_A^{-1} \delta \rho_A)^{(1)} H_A^{(n-k)} (\rho_A^{-1} \delta \rho_A)^{(1)} \rb \rp \cr
& \qquad = \tr\lp \lb H_A^{(k\ge 1)}  \delta \rho_A^{(1)} H_A^{(n-k)} (\rho_A^{-1} \delta \rho_A)^{(1)} \rb \rp  \, ,
\end{align}
which vanishes by cyclicity of the trace.  The remaining terms give
\begin{align} \label{eq:delta2RAraw}
\delta^2 R_A
&= \frac{1}{2} \tr\lp e^{H_A} \delta\rho_A^2 \rp + \sum_{n=1}^\infty \lp \frac{ B_n}{n!}  \rp \tr\lp e^{H_A} \lb H_A^{(n)} \delta\rho_A^{(1)}  \rb \delta\rho_A\rp  \cr
& \qquad \qquad +  \sum_{n=1}^\infty (-1)^n D_{n,0} \tr\lp  \lb \lp e^{H_A}\rp^{(1)}  H_A^{(n)} \delta\rho_A\rb  \delta\rho_A\rp  \, .
\end{align}
Note that when $[H_A,\delta\rho_A]=0$,~\eqref{eq:delta2RAraw} reduces to the term studied in~\cite{Rosenhaus:2014woa}.

We now evaluate the sums in~\eqref{eq:delta2RAraw}.  It is useful to rewrite the first trace as
\begin{align} \label{eq:FirstTrace}
\tr\lp e^{H_A} \lb H_A^{(n)} \delta\rho_A^{(1)}  \rb \delta\rho_A\rp &= \sum_{j=0}^n (-1)^j {n \choose j}\tr\lp e^{H_A}  H_A^{n-j} \delta\rho_AH_A^{j}  \delta\rho_A\rp \cr
&= \sum_{j=0}^n (-1)^j {n \choose j} \left< e^{ H_A}  H_A^{n-j} \delta\rho_AH_A^{j}  \delta\rho_Ae^{ H_A}\right> \cr
&=   \left< \lp e^{H_{A}} \, \delta\rho_A  \rp  (-K)^{n}  \lp \delta\rho_A \, e^{H_{A}} \rp \right> \, ,
\end{align}
where the first line is straightforward to prove by induction. In the second equality we have inserted a factor of $\e^{H_A}\rho_A = 1$ into the trace to convert it to an expectation value with respect to the reduced density matrix $\rho_A$. Since all of the operators inside the expectation value live on $A$, this is equivalent to the expectation value with respect to the full density matrix $\rho = \ket{0}\bra{0}$, which, as in the main text, is denoted by $\ev{\dots}$. The last equality follows from $H_A\ket{0} = H_{A^c}\ket{0}$, the definition $K:= H_{A}-H_{A^c}$, and the binomial expansion.

Similarly, the second trace in~\eqref{eq:delta2RAraw} can be rewritten in the form
\begin{align}\label{eq:SecondTrace}
&\tr\lp  \lb \lp e^{H_A}\rp^{(1)}  H_A^{(n)} \delta\rho_A\rb  \delta\rho_A\rp \cr
&\qquad= \sum_{j=0}^n (-1)^j {n \choose j}\tr\lp H_A^{n-j}  \lb e^{H_A} , \delta\rho_A\rb H_A^{j}  \delta\rho_A\rp \cr
&\qquad= \sum_{j=0}^n (-1)^j {n \choose j} \lp \left<   \lp  e^{H_{A}} \delta\rho_A \rp  H_{A^c}^{n-j}  H_A^{j}  \lp   \delta\rho_A \, e^{H_{A}} \rp \right>  - \left<   \lp e^{H_{A}} \delta\rho_A\rp  H_{A}^{n-j}  H_{A^c}^{j}  \lp   \delta\rho_A \, e^{H_{A}} \rp \right> \rp \cr
&\qquad= \lb (-1)^n - 1\rb  \left<   \lp e^{H_{A}} \delta\rho_A \rp  K^n  \lp  \delta\rho_A \, e^{H_{A}}\rp \right>  \, ,
\end{align}
which clearly vanishes for even $n$.  Inserting~\eqref{eq:FirstTrace} and~\eqref{eq:SecondTrace} into~\eqref{eq:delta2RAraw} and performing the sums gives
\begin{align} \label{eq:delta2RA}
\delta^2 R_A = \left< e^{H_A} \delta\rho_A \lb f(K) - f'(K) \rb  \delta\rho_A \, e^{H_A} \right> \, ,
\end{align}
where $f$ is the smooth function defined in~\eqref{eq:fdef}.  Note that here (and in previous expressions involving correlators) the operators $H_A$, $\delta \rho_A$ should be understood as tensor products with the identity operator on $A^c$.  We computed the above sums using~\eqref{eq:generatingFunction},~\eqref{eq:Dn0} and the trick
\begin{align} \label{eq:sumTrick}
\sum_{n=1}^\infty \frac{B_{n+1}}{n!} (-y)^n
 = -\partial_y  \lp \sum_{n=0}^\infty\frac{B_{n}}{n!} (-y)^{n} - \frac{y}{2} \rp
= - f'(y)  + \frac{1}{2}   \, .
\end{align}

Our manipulations of the infinite sums~\eqref{eq:generatingFunction} and~\eqref{eq:sumTrick} are somewhat formal since each sum converges only for $y\in (-2\pi,2\pi)$ while the eigenvalues of $H_A$ are unbounded.  Nevertheless, since both the initial and final expressions involve only analytic functions of $H_A$, we expect that with appropriate care the final result can be justified for general $H_A$.  The representation of $\ln \rho$ used in e.g. appendix C of \cite{Faulkner:2014jva} may be useful for this purpose.

Note that since $K$ is Hermitian and
\begin{equation}
\label{eq:ffp}
f(y)-f'(y) = \frac{-1+y+e^{-y}}{(1-e^{-y})^2} > 0
\end{equation}
for real $y$, the eigenvalues of $f(K)-f'(K)$ are real and positive.  This fact implies $\delta^2 R_A \ge 0$, with equality if and only if $\delta \rho_A e^{H_A}$ annihilates the vacuum. But since the restriction of $|0\rangle$ to $A$ is non-degenerate, this can occur only if $\delta\rho_A=0$.  Thus~\eqref{eq:delta2RA} is consistent with positivity of the relative entropy.

\bibliographystyle{kp}

\bibliography{FirstLawAtLargeNv17.bbl}

\end{document}